\documentclass[Afour,sageh,times]{sagej}

\usepackage{moreverb,url}
\usepackage[colorlinks,bookmarksopen,bookmarksnumbered,citecolor=red,urlcolor=red]{hyperref}
\usepackage{amssymb}
\usepackage{hyperref}
\usepackage{amsmath}
\usepackage{makecell}
\usepackage{algorithmic}
\usepackage{array}
\usepackage{comment}
\usepackage{textcomp}
\usepackage{stfloats}
\usepackage[ruled]{algorithm2e}
\usepackage{url}
\usepackage{verbatim}
\usepackage{multirow}
\usepackage{graphicx}
\usepackage{xcolor}
\usepackage{array}
\usepackage{soul}
\usepackage{balance}
\usepackage{adjustbox}
\usepackage{booktabs}
\usepackage{arydshln}

\newcommand{\myhang}[2]{%
  \par\begingroup
  \noindent
  \hangindent=#1
  \hangafter=0
  #2\par
  \endgroup
}

\newcommand\BibTeX{{\rmfamily B\kern-.05em \textsc{i\kern-.025em b}\kern-.08em
T\kern-.1667em\lower.7ex\hbox{E}\kern-.125emX}}

\def\volumeyear{2016}

\begin{document}

\runninghead{Yuequan Bao and Xing Li}

\title{SHM-Agents: A Generalist-Specialist Integrated Agent System for Structural Health Monitoring}

\author{Yuequan Bao\affilnum{1}\affilnum{2}\affilnum{3}, Xing Li\affilnum{3},  Huabin Sun\affilnum{3}, Dawei Liu\affilnum{3},  Yuxuan Tian\affilnum{3},  Haiyang Hu\affilnum{3}}

\affiliation{\affilnum{1}Key Lab of Smart Prevention and Mitigation of Civil Engineering Disasters of the Ministry of Industry and Information Technology, Harbin Institute of Technology, Harbin\\
\affilnum{2}Key Lab of Structures Dynamic Behavior and Control of the Ministry of Education, Harbin Institute of Technology, Harbin\\
\affilnum{3}School of Civil Engineering, Harbin Institute of Technology, Harbin}

\corrauth{Yuequan Bao, chool of Civil Engineering,
Harbin Institute of Technology, 73Huanghe Road,
Harbin 150001, China.}

\email{baoyuequan@hit.edu.cn.}

\begin{abstract}
Artificial intelligence is increasingly used to simplify complex tasks. In engineering applications of structural health monitoring (SHM), existing specialized algorithms, while effective, often face high implementation barriers, limited interoperability and complex training procedures. To overcome these challenges, this paper proposes SHM-Agents, a generalist–specialist agent system that integrates the reasoning and planning abilities of large language models with the problem-solving strengths of specialized algorithms. SHM-Agents enables end-to-end execution of single and combined SHM tasks via natural language, supports deep learning pre-training to simplify deployment and allows flexible expansion through a modular design. Experiments on a long-span cable-stayed bridge show that SHM-Agents can accurately and efficiently perform diverse SHM tasks, including data anomaly diagnosis and recovery, signal processing, statistical analysis, modal identification, damage identification, finite element model updating, vehicle load modeling, response calculation, reliability assessment, fatigue estimation and bridge knowledge Q\&A.
\end{abstract}


\maketitle
\section{Introduction}

In recent years, rapid advances in artificial intelligence have led to the emergence of numerous large language models, including GPT(\cite{art6}), DeepSeek(\cite{art7}), LLaMA(\cite{art8}) and Gemini(\cite{art9}), which have demonstrated remarkable capabilities in natural language understanding and task execution. Consequently, the application of artificial intelligence to complex engineering problems has become a major research focus because of its potential to streamline workflows and improve efficiency. However, owing to the inherent limitations of large language models, their outputs are predominantly text-based, which prevents them from independently completing complex tasks. To overcome this limitation, researchers have proposed large language model-based agents (LLM agents). 

\begin{figure}[!t]
	\centering
	\includegraphics[width=0.98\columnwidth]{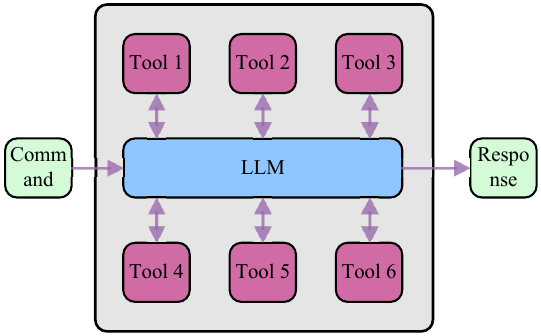}
	\caption{{ The Architecture of LLM Agents.}}
	\label{fig2}
\end{figure}

\begin{figure*}[!t]
	\centering
	\includegraphics[width=0.98\textwidth]{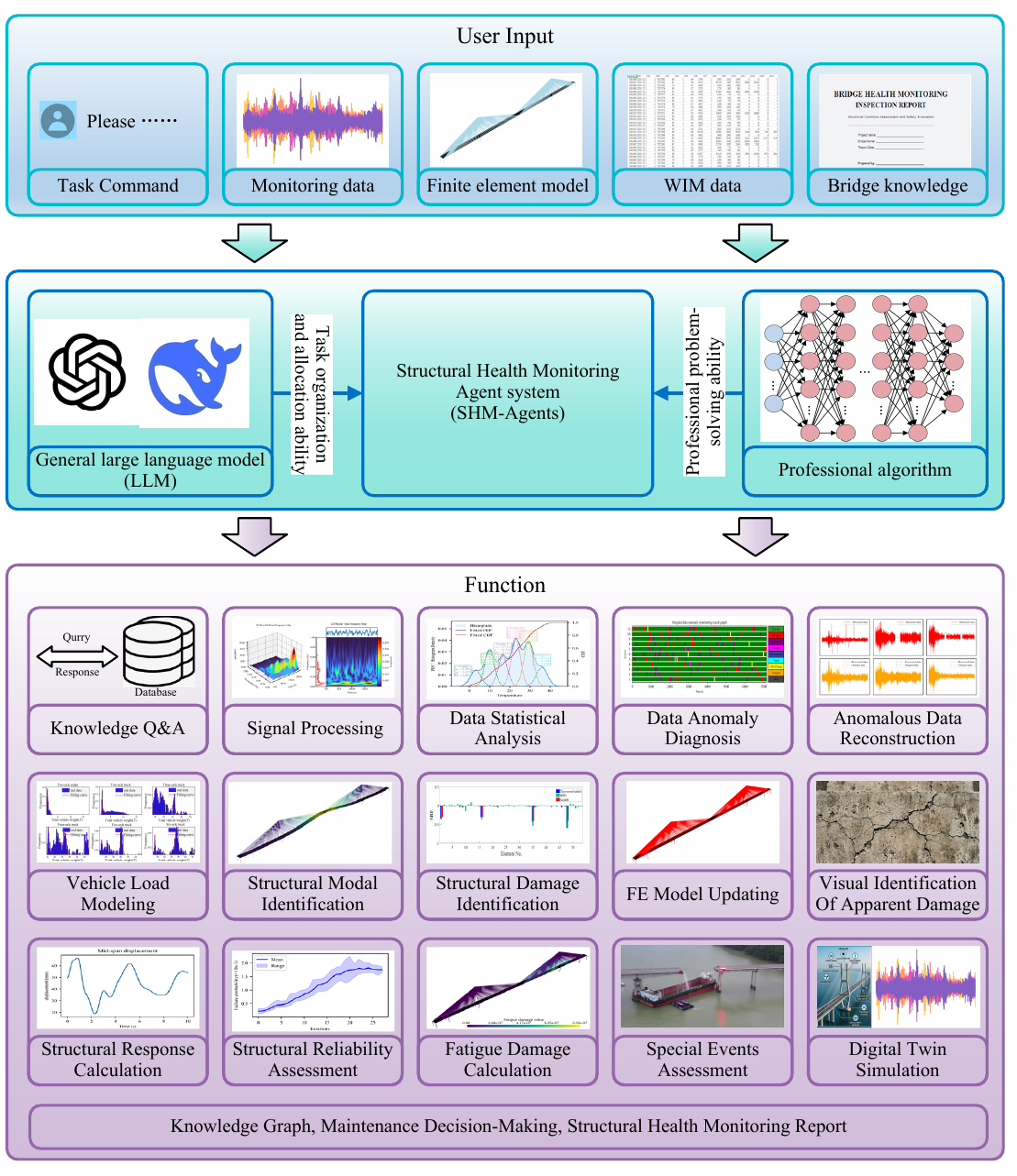}
	\caption{{ The functions of the SHM-Agents.}}
	\label{fig1}
\end{figure*}

LLM agents are systems designed to enable large language models (LLMs) to reason through problems, formulate solutions and execute plans using a suite of tools. Their primary function is to serve as intermediaries between LLMs and APIs or other external tools, allowing LLMs to leverage these resources to accomplish complex tasks, as illustrated in Figure \ref{fig2}. To streamline agent development, researchers have introduced numerous open-source frameworks (\cite{art24}, \cite{art46,art47,art48,art49,art50,art51,art52,art53,art54}), including ChatDev (\cite{art46}), Auto-GPT (\cite{art47}), MetaGPT (\cite{art49}), LangChain (\cite{art50}) and LangGraph (\cite{art48}). The introduction of these frameworks has substantially reduced the complexity of agent construction, thereby accelerating their adoption across various domains (\cite{art10,art11,art12,art13,art14,art15,art16,art17,art18,art19,art20,art21,art22,art23}), such as aerospace, biology, chemistry, law and finance.

\cite{art14} proposed a modular multi-agent LLM framework for scientific tasks, enabling the collaborative execution of complex workflows through the flexible integration of customized Python functions. Building on this framework, they developed DynaMate, a system that automates the generation, execution and analysis of molecular simulations. \cite{art23} proposed an LLM-based LRC agent for human–robot collaboration that integrates chain-of-thought reasoning, few-shot learning and contextual memory. Experiments in aerospace wire harness assembly demonstrated that the agent can accurately interpret instructions, plan tasks effectively and support flexible and efficient robotic operations. \cite{art11} introduced Agent4Vul, a multimodal LLM-agent framework for smart contract vulnerability detection that incorporates Commentator and Vectorizer agents within a multimodal learning architecture. The results showed that Agent4Vul outperformed a broad range of baseline methods and surpassed existing state-of-the-art AI models. \cite{art21} investigated digital twins constructed using LLM agents to simulate human multi-turn interactions with Amazon Rufus. Their findings indicated that these digital twins reproduced human-like behavioral patterns while enabling scalable evaluation of agentic AI systems. \cite{art12} proposed MEDRAG, a medical retrieval-augmented generation framework that integrates medical corpora, retrieval modules and backbone LLMs for medical question answering. Experimental results showed that MEDRAG improved answer accuracy, reduced hallucinations and outdated information and provided practical guidance for the design of medical RAG systems.

In the field of SHM, establishing an effective monitoring system is essential for tracking structural operating conditions, predicting potential damage and deformation trends and ensuring structural safety (\cite{art1,art2,art3,art4}). Based on monitoring data, image, traffic information and finite element model, the SHM system performs structural analysis using specialized algorithms(\cite{art25,art26,art27,art28,art29,art30,art31,art32,art33,art34,art35,art36,art37,art38,art39,art40,art41,art42,art43,art44,art45}), including modal identification, damage detection, reliability assessment and fatigue damage evaluation, to determine the health condition of the structure and provide recommendations for operation and maintenance. In recent years, the rapid advancement of deep learning has led to significant progress in specialized algorithms across a wide range of applications, thereby substantially improving the performance of SHM systems. However, existing SHM algorithms still suffer from limited integration, fragmented workflows and high barriers to adoption. Users are required not only to understand and operate multiple independent algorithmic modules with incompatible interfaces, but also to invest considerable time in data preparation, model training, inference deployment and manual workflow coordination. Consequently, these systems remain inefficient and difficult to use in practice.

To address this issue, this paper proposes SHM-Agents, a generalist-specialist integrated agent system that combines the reasoning and planning capabilities of general large language models with the problem-solving capabilities of specialized algorithms. The system supports a wide range of natural language-driven functions, including data anomaly diagnosis, anomalous data recovery, signal processing, statistical analysis, structural mode identification, structural damage identification, finite element model updating, structural response calculation, structural reliability assessment, structural fatigue damage estimation, vehicle load modeling and bridge knowledge Q\&A. These functions can be used either individually or in combination. The specific contributions of this paper are as follows.
\begin{itemize}
\item{This paper presents the architecture and development of a system called SHM-Agents. The system employs a large language model as its intelligent core and integrates a range of specialized algorithms for SHM, demonstrating strong capabilities in organizing, allocating and executing complex tasks. It performs tasks according to an “architecture–allocation–execution–summary” workflow and ensures successful completion through a self-feedback regulation mechanism.}
\item{The SHM-Agents enables end-to-end execution of individual or combined tasks in the field of SHM through natural language interaction. Supported by configuration and pre-training strategies, the system can complete the deep model pre-training process without human intervention, thereby significantly simplifying the use of specialized algorithms.}
\item{The SHM-Agents introduces an Algorithm Module to manage diverse specialized algorithms and ensure their correct invocation during task execution. It also includes a Data Module to manage variables generated during execution and to ensure their accurate transfer among SHM algorithms.}
\item The SHM-Agents integrates a wide range of algorithms in a modular manner for SHM, enabling the execution of both individual and combined tasks. Moreover, SHM-Agents is highly scalable, allowing users to rapidly add new agent nodes to address additional tasks.
\end{itemize}

The remainder of this paper is organized as follows. Section 2 details the structure and methodologies of the proposed agent system. Section 3 illustrates practical application examples of this agent system. Section 4 lists the known limitations of SHM-Agents. Section 5 concludes the paper.

\section{Methodology}
\label{s2}

This section will introduce the generalist-specialist integrated agent system, SHM-Agents, proposed in this paper and provide a detailed explanation of its architectural design and usage methods.

\subsection{Overview}

To enhance the convenience of deep learning algorithms for solving specific problems in SHM, this paper proposes a LLM agents system called SHM-Agents. SHM-Agents leverage natural language commands for control and can intelligently invoke one or more specialized algorithms to complete specific tasks. 

\begin{figure*}[!t]
	\centering
	\includegraphics[width=0.98\textwidth]{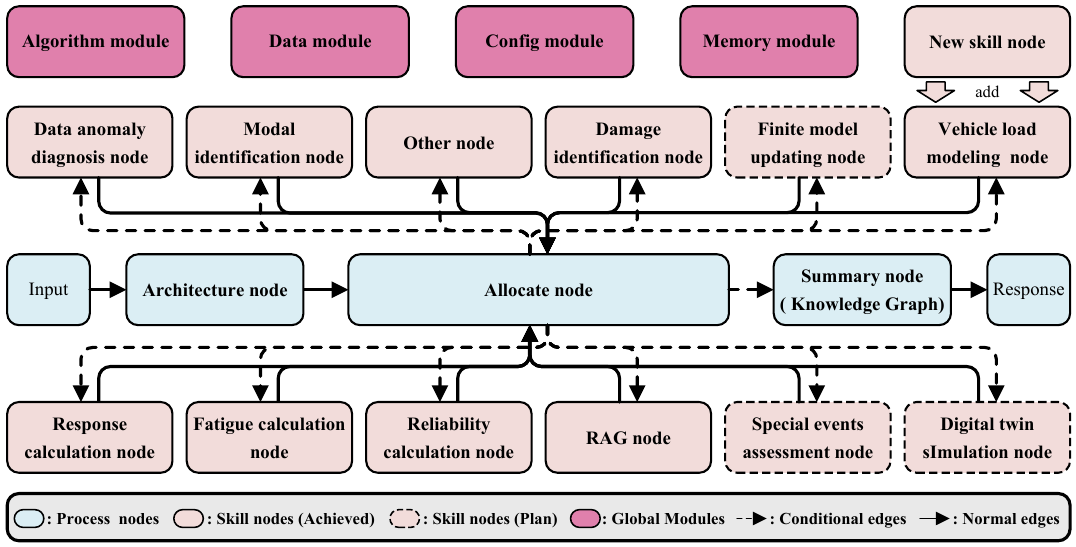}
	\caption{{ The Architecture of SHM-Agents. SHM-Agents comprise three key components: process agent nodes, skill agent nodes and global modules. The process agent nodes oversee task planning, allocation, summarization and output. The skill agent nodes are tasked with executing specific SHM algorithms. The global modules manages operational information, including memory, data and configurations. Conditional edges refer to the edges that determine the next routing node based on program decisions.}}
	\label{fig3}
\end{figure*}

The architecture of SHM-Agents is illustrated in Figure \ref{fig3} and features a modular design consisting of process agent nodes, skill agent nodes and global modules. The process agent nodes are divided into three types: the Architecture node, which generates task execution plans; the Allocate node, which evaluates and assigns tasks; and the Summary node, which summarizes operational processes and returns the results. The skill agent nodes perform specific SHM tasks and their number can be adjusted, allowing users to add new skill nodes as needed. The global modules comprise several submodules: a memory module that manages the context of agent dialogues, thereby providing the system with memory capabilities; an algorithm module that stores the input and output specifications of specialized algorithms to ensure their accurate invocation; a data module that retains user input and operationally generated data, thereby ensuring accurate data transmission across nodes; and a configuration module that stores finite element models, optimal parameters for specialized deep learning algorithms and LLM configuration parameters. Together, these modules enable the agent system to perform SHM tasks accurately and efficiently. 

\subsection{Global modules}

The global modules are essential components that store information integral to the operation of the SHM-Agents system. These encompass the algorithm module, data module, config module and memory module. The following sections will detail the rationale behind their construction, the methods employed in building them and their respective functionalities. 

\textbf{Algorithm Module:} To ensure that the LLM comprehends the functions, input requirements and output formats of various SHM algorithms, SHM-Agents developed an algorithm module. This module is implemented using Python dictionaries, with each specialized algorithm represented by a dictionary containing key-value pairs for the algorithm's function, input requirements, input examples and output examples.

\textbf{Data Module:} In SHM tasks, multiple specialized algorithms often need to collaborate. However, the variations in input data formats and shapes required by each algorithm can hinder direct data transfer between skill agent nodes. To address this challenge, SHM-Agents developed a data module comprising methods for data management, reading, querying, saving and transformation. The data management method, implemented using Python dictionaries, records variable names (memory addresses), data descriptions, storage locations, types and shapes. The methods for data reading, querying, saving and transformation are implemented through four simple LLM Agents. These methods can programmatically write and execute Python scripts as needed to facilitate data reading, detailed querying, local storage and format transformation.

\textbf{Config Module:} SHM-Agents aims to establish an intelligent monitoring system for any bridge. However, the configuration information (such as the optimal parameters for specialized deep algorithms, database details and finite element model) varies across different bridges and generating this information typically requires substantial time. To streamline the generation and management of configuration information for various bridges, SHM-Agents developed a config module. This module comprises a configuration management component and a configuration generation component. Using Python dictionaries, the configuration management component stores comprehensive configuration details for each bridge. Users simply select the configuration for a specific bridge to execute the health monitoring task for that bridge. The configuration generation component is responsible for producing the configuration data for each bridge. It uses algorithms such as deep model pre-training, database pre-construction and finite element model preprocessing to transform user-provided configuration data into the configuration required for agent operation, storing this information locally.

\textbf{Memory Module:} In the operation of the SHM-Agents, multiple agent nodes, such as Allocate node and Summary node, must integrate previous dialogue information to make informed judgments and responses. To address this, SHM-Agents have developed a memory module. This module is specifically built using Python's TypedDict method, allowing for the storage of contextual information throughout the dialogue process. Consequently, this endows the agent system with the capability to retain and utilize conversational memory.

\subsection{Skill agent nodes}

In the SHM-Agents system, skill-agent nodes are designed to perform specific tasks. These nodes provide excellent scalability, allowing users to add new functionalities with minimal coding effort. At present, the system includes 8 agent nodes dedicated to SHM, each specializing in a particular task. The functionality of these agents is implemented using corresponding algorithms. Table \ref{tab1} provides a comprehensive overview of the capabilities of each skill-agent node and its associated tools. As shown in \ref{fig2}, each node functions as a basic agent.

\begin{table*}[!t]
	\begin{center}
		\caption{\small Capabilities and tools of skill agents.}
		\label{tab1}
		\begin{tabular}{c c c}
			\toprule
			\textbf{Node name}& \textbf{Node function}  & \textbf{Tools}  \\
			\midrule
			Data anomaly diagnosis node& \makecell{Anomaly detection and abnormal data  \\ recovery for monitoring data.}   & \makecell{Data anomaly monitoring algorithm \\ (\cite{art31}) \\ Abnormal data recovery algorithm} \\ \hline

			Modal identification node& \makecell{Identification of structural modal \\ parameters based on monitoring data.}& \makecell{Modal identification algorithm \\ (\cite{art33})}\\ \hline


			Vehicle load modeling node & \makecell{Modeling vehicle loads based on wim\\ data and generating random traffic flows.} & \makecell{Vehicle load modeling algorithm \\ (\cite{art38})} \\ \hline


			Response calculation node & \makecell{Calculate the response of structure under\\static/dynamic loads (vehicle loads).} & \makecell{A series of functional functions \\ based on PyAnsys} \\ \hline

			Fatigue calculation node & \makecell{Calculate the fatigue damage of structures \\  under long-term vehicle loads.} & \makecell{Rapid calculation algorithm\\ for fatigue damage \\ (\cite{art41})} \\ \hline

			Reliability calculation node & \makecell{Rapid calculation of structural reliability \\ under specified uncertainties.} & \makecell{Rapid calculation algorithm \\ for structural reliability \\ (\cite{art40}) } \\ \hline

			Rag node & \makecell{Perform Q\&A operations  on bridge \\ information based on input files.} & Rag knowledge Q\&A method \\ \hline

			Other node & \makecell{Writing and debugging Python code \\ to accomplish simple tasks in \\ SHM, such as \\ Fourier transform, wavelet transform and \\ probability distribution fitting.} & \makecell{Code writing tool\\ Code execution tool} \\

			\bottomrule
\end{tabular}%
	\end{center}
\end{table*}

\subsection{Process agent nodes}

The process agent nodes constitute the central framework of the SHM-Agents system, comprising the Architecture node, Summary node and Allocate node. They are tasked with planning assignments, summarizing and outputting agent execution results and organizing skill agents to execute associated tasks. The subsequent sections will elucidate their construction methods and functionalities, while also detailing the process through which SHM-Agents execute specific tasks.

\textbf{Architecture node:} In practical health monitoring tasks, it is often necessary to orchestrate multiple specialized agent nodes sequentially to accomplish a specific objective. For example, in structural reliability analysis using monitoring data as input, the process begins by applying data anomaly detection and recovery algorithms to identify and restore anomalous data. Next, structural modal parameters are identified from the recovered data to obtain accurate modal estimates. These modal parameters are then used to update the finite element model. Finally, the updated model is employed to efficiently evaluate structural reliability and estimate the probability of structural failure under real operating conditions. If these tasks are not systematically planned and a large language model (LLM) is used directly to perform them, its performance is likely to be suboptimal. To address this issue, SHM-Agents introduces an Architecture node that plans the entire workflow before task execution, thereby ensuring a systematic completion process. The Architecture node performs this function through customized prompts designed for the LLM. By incorporating algorithmic details, data characteristics and task execution examples into these prompts, it enables the LLM to understand the task requirements and the functions of each specialized agent node, thereby formulating an effective execution strategy.

\begin{figure}[!t]
	\centering
	\includegraphics[width=0.98\columnwidth]{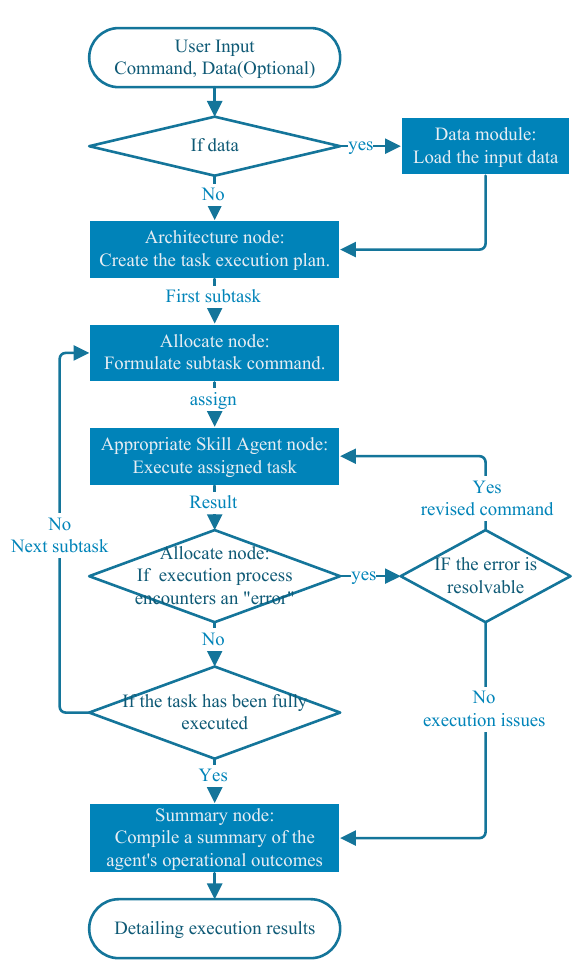}
	\caption{{ The flowchart of SHM-Agents executing specific tasks.}}
	\label{fig4}
\end{figure}

\textbf{Summary node:} The Summary node in SHM-Agents is designed to encapsulate the dialogue involved in the task execution process, produce a comprehensive runtime summary and deliver feedback to the user. This summary encompasses the execution flow, outcomes, images and the storage locations for data.

\textbf{Allocate node:} After the Architecture node generates the task execution plan, execution begins. Because multiple skill agent nodes must be invoked sequentially, a dedicated manager is required to monitor execution and assign tasks appropriately. In SHM-Agents, this role is performed by the Allocate node. First, the Allocate node analyzes the task workflow, generates a detailed execution command for the next task step and sends it to the corresponding skill agent node. After the command is executed, the skill agent returns the result to the Allocate node. The Allocate node then evaluates the result. If the skill agent encounters a correctable error during execution, the Allocate node regenerates the command to ensure proper execution. If no error is detected, the workflow proceeds to the next task step. If an error cannot be corrected, the Allocate node reports it directly to the user. Once all subtasks have been completed, the Allocate node terminates the execution process and transfers control to the Summary node for final output. This distinctive “allocate-execute-feedback-reallocate” mechanism enables automatic error correction during execution, thereby ensuring the smooth operation of SHM-Agents and preventing program crashes caused by execution errors.

The pseudocode for SHM-Agents performing specific tasks is shown as Algorithm 1 and the flowchart is illustrated in Figure \ref{fig4}.

\noindent\rule{\linewidth}{1.5pt}

\noindent{\textbf{  Algorithm 1.} The procedure of SHM-Agents executing specific tasks.}

\noindent\rule{\linewidth}{1pt}

    \KwIn{\\
	\myhang{2em}{A natural language command} \\
	\myhang{2em}{Monitoring/Detection Data or other Data (Optional)}
	}	
    \KwOut{\\
	\myhang{2em}{A natural language segment detailing execution results, including images and the storage locations data.\\}}

	\vspace{1em}

	\textbf{1.preprocessing}

	\textbf{IF} the input includes data \textbf{THEN}

	\myhang{2em}{Data module: Load the input data into the data module.}

	\textbf{ENDIF}

	\vspace{2em}
	\textbf{2. Task Planning}

	Architecture node: Create the task execution plan based on the input.

	\vspace{2em}
	\textbf{3. Task Execution}

	\myhang{2em}{\textbf{For} each subtask \textbf{in} the task plan:}
	
	\myhang{3em}{Allocate node: Formulate subtask command and assign it to the appropriate skill agent node.}
	
	\myhang{3em}{Skill Agent node: Execute assigned task and log results within the data module.}

	\myhang{3em}{\textbf{IF} the skill agent execution process encounters an "error" \textbf{THEN}}

	\myhang{4em}{\textbf{IF} the error is resolvable \textbf{THEN}}

	\myhang{4em}{Generate a revised command and forward it to the skill agent node.}
	
	\myhang{3em}{\textbf{ELSE}}
	
	\myhang{4em}{Notify user of execution issues. }

	\myhang{4em}{\textbf{EXIT LOOP}}

	\myhang{3em}{\textbf{ENDIF}}

	\myhang{2em}{\textbf{ENDIF}}

	\myhang{2em}{\textbf{END FOR}}

	\vspace{1em}
	\textbf{4. Results Summary and Output}

	\myhang{2em}{Summary node: Compile a summary of the agent's operational outcomes and communicate these results to the user.}

\noindent\rule{\linewidth}{1.5pt}

\subsection{User Interface and Instructions}

\begin{figure*}[!t]
	\centering
	\includegraphics[width=0.98\textwidth]{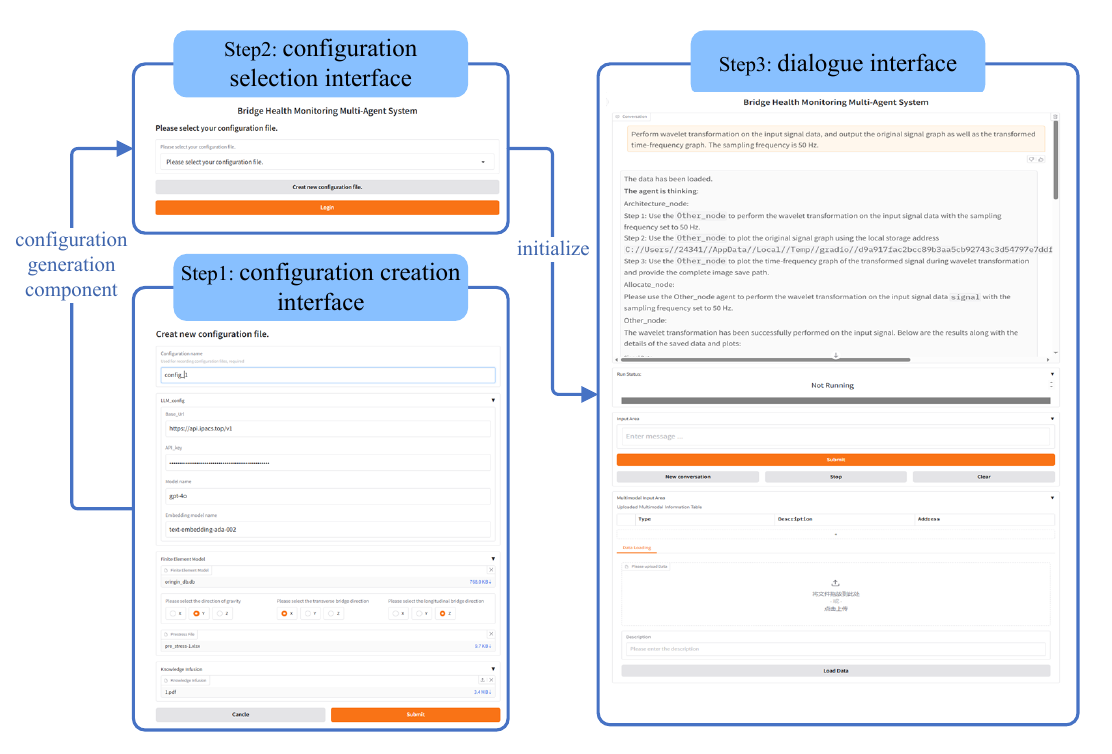}
	\caption{{ The user interface and Instructions of SHM-Agents.}}
	\label{fig5}
\end{figure*}

To improve user interaction with the SHM-Agents system, a user interface has been developed. As shown in Figure \ref{fig5}, the interface consists of three main components: a configuration creation interface, a configuration selection interface and a dialogue interface. In the configuration creation interface, users are required to upload the bridge finite element model, training data, bridge information file and LLM configuration file. SHM-Agents then activates the configuration generation module to create a new configuration. Through the configuration selection interface, users can choose a configuration and log in, after which SHM-Agents generates the corresponding dialogue interface. In this dialogue interface, users can enter commands in the input box or upload data and images through the multimodal box and then submit them to SHM-Agents for processing. The processing steps and results are presented in a conversational format within the dialogue interface.

\section{Experiments}
\label{s3}

This section will demonstrate the application of SHM-Agents in various tasks of real-world SHM.

\subsection{Configuration and Preprocessin}

\begin{figure*}[!t]
	\centering
	\includegraphics[width=0.98\textwidth]{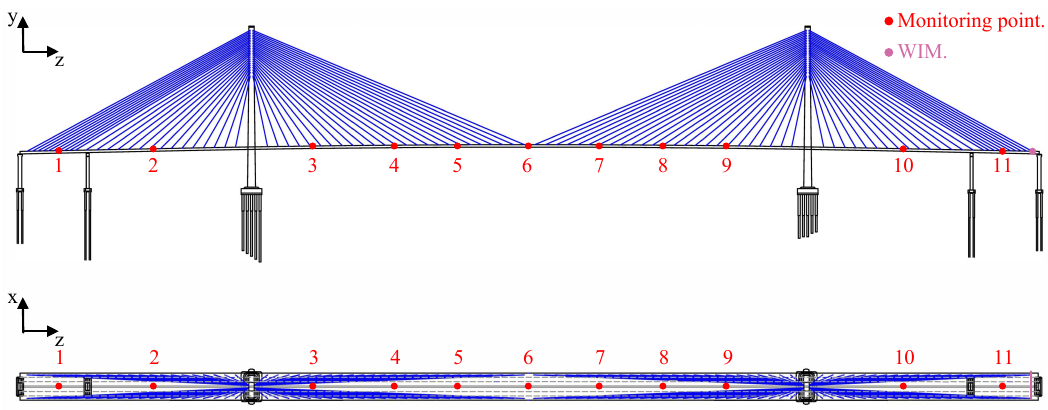}
	\caption{{Diagram of the structure of a cable-stayed bridge.}}
	\label{fig6}
\end{figure*}

An example of applying SHM-Agents is provided by a long-span cable-stayed bridge in China. Figure \ref{fig6} presents a structural diagram of this bridge. It is a dual-direction, six-lane highway bridge, designed to accommodate a Highway Level I vehicle load. Eleven acceleration sensors are installed along the y-axis of the bridge, which collect acceleration data at a frequency of 50Hz. Moreover, a WIM system is installed to gather information on passing vehicles.

\subsubsection{Configuration}

\begin{figure}[!t]
	\centering
	\includegraphics[width=0.98\columnwidth]{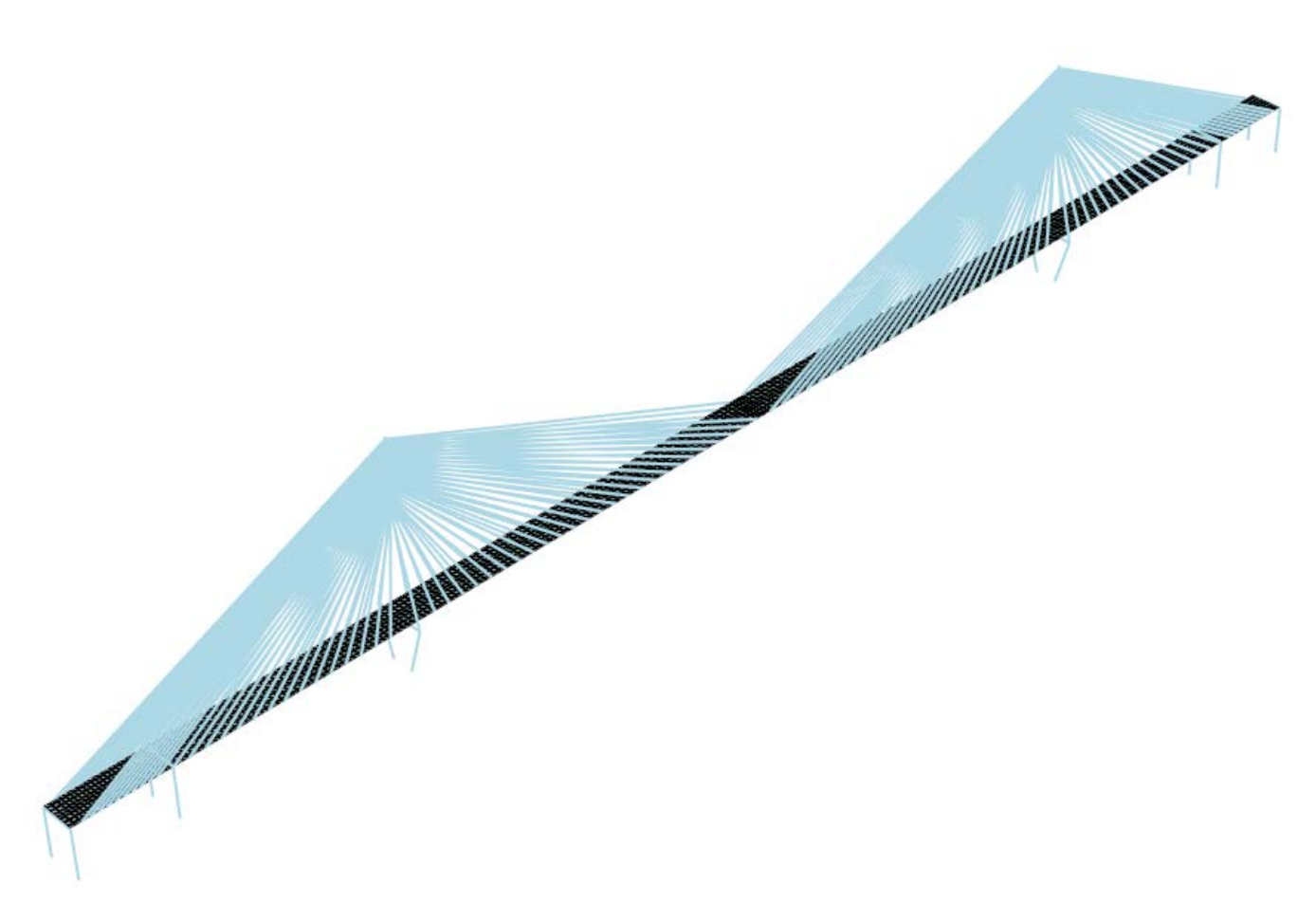}
	\caption{{The finite element model of the cable-stayed bridge.}}
	\label{fig7}
\end{figure}

In order to use SHM-Agents on the aforementioned cable-stayed bridge, a configuration file needs to be created first. The configuration file of SHM-Agents consists of the following four sections.
\begin{enumerate}
    \item \textbf{LLM:} In this case, the chat model used is GPT-4o and the embedding model used is text-embedding-ada-002. Both models are called via API.

    \item \textbf{Finite:} In this case, as illustrated in Figure \ref{fig7}, a finite element model of the cable-stayed bridge was developed using Ansys. Notably, to enable SHM-Agents to apply lane loads effectively, it is essential to predefine lane components within the finite element model.

    \item \textbf{Vehicle load modeling:} In this case, WIM data is added to the SHM-Agents configuration for modeling vehicle loads.

    \item \textbf{Bridge information:} In this case, bridge information, encompassing construction details and maintenance inspection reports, is incorporated into the SHM-Agents configuration in PDF format.
\end{enumerate}

\subsubsection{Preprocessing}

After uploading the configuration parameters, the configuration generation module will perform preprocessing, including finite element model processing, bridge information encoding, vehicle load modeling, data anomaly diagnosis model training, fatigue damage calculation model training and structural damage identification model training. Once the configuration generation module finishes running, the user interface will display "Configuration Created Successfully." Subsequently, users can load this configuration to perform various SHM tasks.

\subsection{Dialogue example}

This section will introduce several complete processes of using SHM-Agents to conduct dialogues for completing specific SHM tasks.

\subsubsection{Data anomaly diagnosis and reconstruct}

The complete dialogue on data anomaly diagnosis and reconstruct based on SHM-Agents is detailed in Dialogue 1.

\begin{figure*}[!t]
	\centering
	\includegraphics[width=0.98\textwidth]{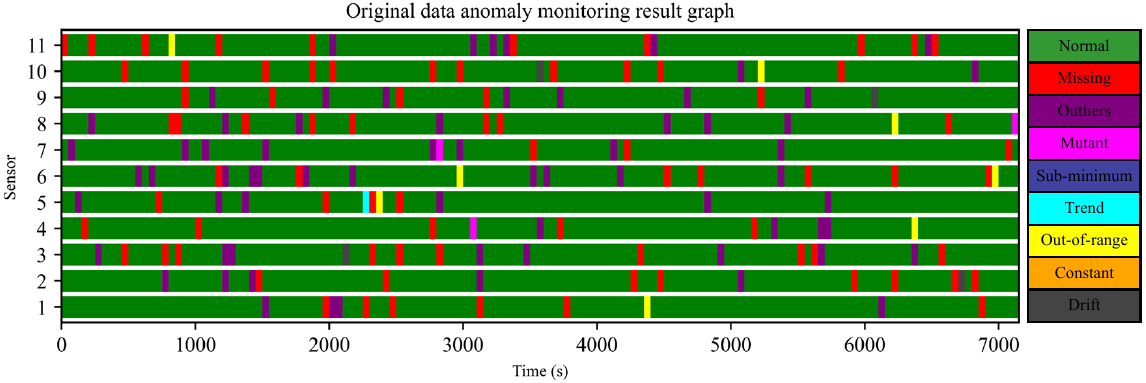}
	\caption{{Anomaly detection results for original data. In this image, the color green indicates normal data, whereas the other colors correspond to various types of anomalies.}}
	\label{fig8}
\end{figure*}

\begin{figure*}[!t]
	\centering
	\includegraphics[width=0.98\textwidth]{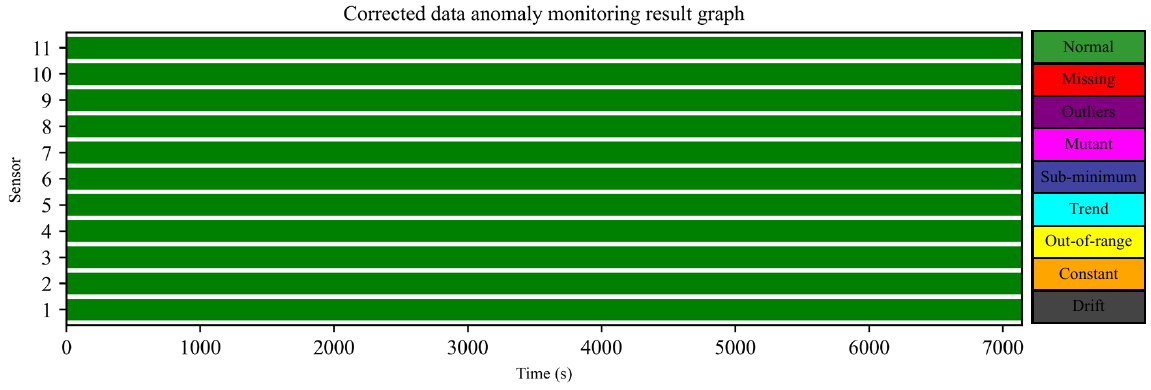}
	\caption{{Anomaly detection results for reconstructed data.}}
	\label{fig9}
\end{figure*}

\begin{figure*}[!t]
	\centering
	\includegraphics[width=0.98\textwidth]{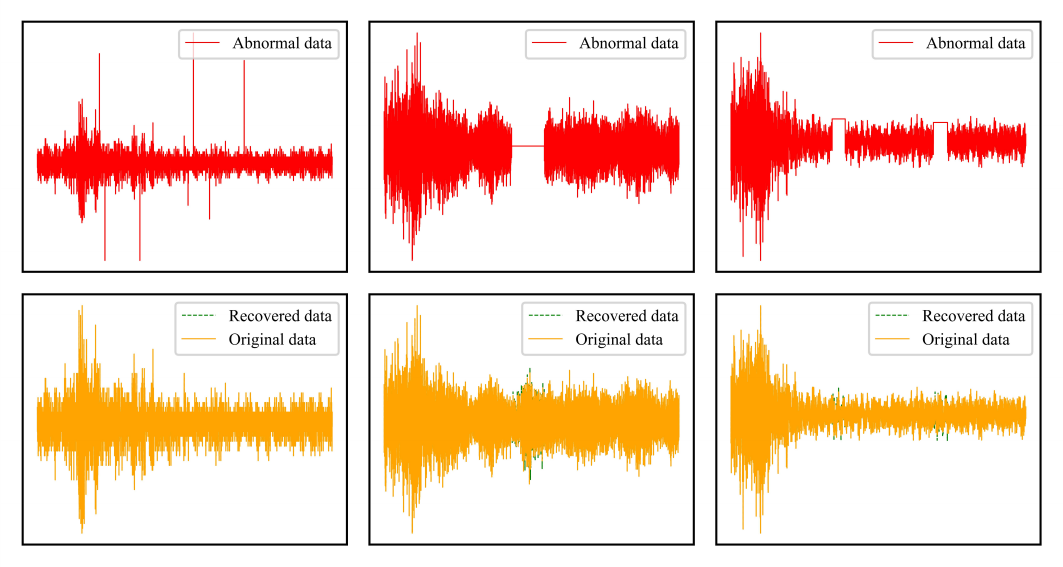}
	\caption{{Original anomaly data and its data recovery results.}}
	\label{fig10}
\end{figure*}

\noindent\rule{\linewidth}{1.5pt}

\noindent{\textbf{  Dialogue 1.} Data anomaly diagnosis and reconstruction.}

\noindent\rule{\linewidth}{1pt}

\textbf{Input:}

\myhang{1em}{1. Command:}
\myhang{2em}{Please perform anomaly diagnosis on the input monitoring data and reconstruct the anomalous data. The time interval is set to 1 minute.}
\myhang{1em}{2. Input data: }
\myhang{2em}{Monitoring data: Monitoring data from acceleration sensors.}

\vspace{1em}
\textbf{Thinking:}

1.Architecture node:

\myhang{2em}{Step 1: Use the Data Anomaly Diagnosis node agent to conduct long-term data anomaly detection on the input \texttt{monitoring\_data} with a time interval set at 60 seconds.}

2.Allocate node:

\myhang{2em}{Please use the Data Anomaly Diagnosis node agent to conduct long-term data anomaly detection on the input \texttt{monitoring\_data} with the time interval set at 60 seconds.}

3.Data Anomaly Diagnosis node:

\myhang{2em}{The long-term data anomaly detection task has been successfully executed. Below are the results:}
\myhang{2em}{1.	Original Data Anomaly Diagnosis Result: The graph is show in Figure \ref{fig8}.}
\myhang{2em}{2.	Reconstruction Data: Abnormal data has been reconstructed successfully. The reconstructed data is stored in the variable \texttt{Effective\_monitoring\_data}.}
\myhang{2em}{3.	Reconstruction Data Anomaly Diagnosis Result: The graph is show in Figure \ref{fig9}.
}

\vspace{1em}

\textbf{Ouput:}
\myhang{1em}{The long-term data anomaly detection task has been successfully executed. Below are the results:}
\myhang{1em}{1.	Original Data Anomaly Diagnosis Result: The graph is show in Figure \ref{fig8}.}
\myhang{1em}{2.	Reconstruction Data: Abnormal data has been reconstructed successfully. The reconstructed data is stored in the variable \texttt{Effective\_monitoring\_data}.}
\myhang{1em}{3.	Reconstruction Data Anomaly Diagnosis Result: The graph is show in Figure \ref{fig9}.
}

\noindent\rule{\linewidth}{1.5pt}

As shown in Dialogue 1, SHM-Agents is driven by natural language and can load data, execute specified tasks and generate natural-language text and images containing task-completion information. In the data anomaly diagnosis task, SHM-Agents first segments the data according to the time interval specified by the user. It then performs anomaly diagnosis on each segment. For segments identified as anomalous, SHM-Agents reconstructs the data by incorporating alignment information from the preceding and subsequent time periods. After reconstruction, SHM-Agents performs anomaly diagnosis again on the reconstructed data to assess the reconstruction quality. Finally, the system saves the reconstructed data as a new variable and returns it to the user. To evaluate the model’s reconstruction performance, this study manually introduces anomalies into a subset of the data and compares the recovered images with the corresponding ground-truth images, as shown in Figure \ref{fig10}.

\subsubsection{Simple SHM task}

To address tasks beyond the scope of specialized models, SHM-Agents developed an intelligent agent called "Other Node." This agent leverages the capabilities of large language models to generate and execute code for accomplishing specific tasks. Its effectiveness has been demonstrated through two example tasks: signal analysis and probability distribution fitting.

The complete dialogue for performing wavelet transform using SHM-Agents is detailed in dialogue 2.

\begin{figure*}[!t]
	\centering
	\includegraphics[width=0.98\textwidth]{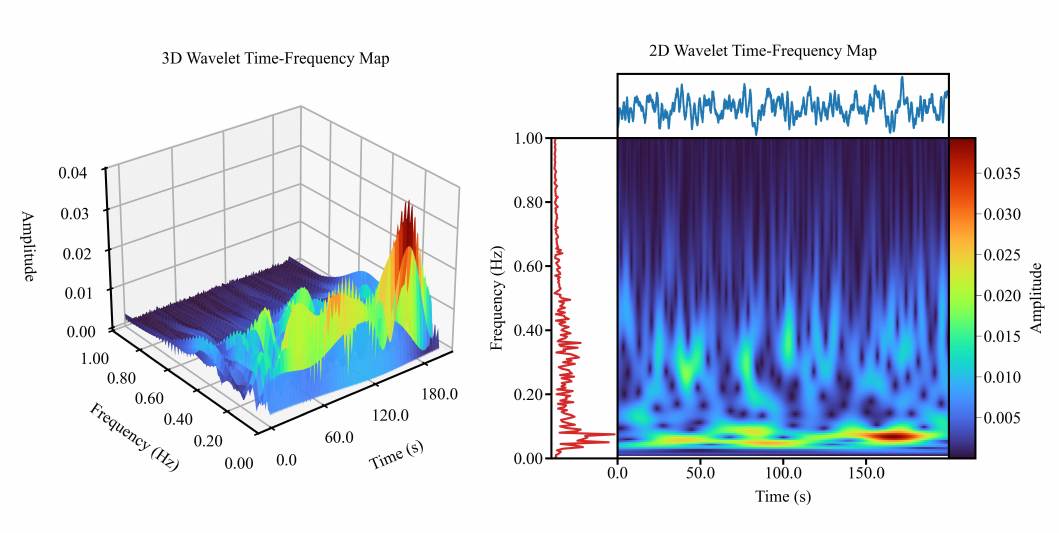}
	\caption{{Wavelet transform result.}}
	\label{fig11}
\end{figure*}

\noindent\rule{\linewidth}{1.5pt}

\noindent{\textbf{  Dialogue 2.} Wavelet transform.}

\noindent\rule{\linewidth}{1pt}

\textbf{Input:}

\myhang{1em}{1.	Command:}

\myhang{2em}{Please perform a Fourier transform and a wavelet transform on the input signal (with a sampling frequency of 50 Hz), and meet the following requirements:}
\myhang{2em}{(1). Apply both the Fourier transform and the wavelet transform to the input signal, and extract only the 0–1 Hz frequency components for plotting.}
\myhang{2em}{(2). Use a single canvas with left and right subplots, with a total canvas width of 7.06. The left subplot should show a 3D time-frequency plot of the wavelet transform.}
\myhang{2em}{(3). The right side should contain:}
\myhang{3em}{- a 2D time-frequency plot of the wavelet transform,}
\myhang{3em}{- the original data plot,}
\myhang{3em}{- and the Fourier transform spectrum plot.}
\myhang{3em}{The original data plot should be placed directly above the 2D time-frequency plot, with their time axes aligned. The original data plot should have no time-axis ticks, labels, or title. The Fourier transform spectrum plot should be placed vertically on the left side of the 2D time-frequency plot, with their frequency axes aligned. The frequency axis of the 2D time-frequency plot should have no axis ticks, labels, or title. Only one title is needed for the three plots on the right. Please note that the colorbar on the right side of the 2D time-frequency plot may cause misalignment with the time axis of the original data plot.}

\myhang{1em}{2. Input data:}
\myhang{2em}{Monitoring data: Monitoring data from an acceleration sensor.}

\vspace{1em}

\textbf{Thinking:}

1. Architecture node: 
\myhang{2em}{Step 1: Use the Other node agent to perform both a Fourier transform and an appropriate wavelet transform on the input signal, which is a numpy.ndarray of shape (10000,) sampled at 50 Hz. Extract and plot only the 0–1 Hz frequency components. Create a single figure with a total width of 7.06, consisting of a 3D wavelet time–frequency subplot on the left and, on the right, a composite layout in which the original data plot is placed directly above the 2D wavelet time–frequency plot with aligned time axes, while the Fourier spectrum is positioned vertically to the left of the 2D wavelet plot with aligned frequency axes. Ensure that the original data plot has neither time-axis tick labels nor a title, and that the 2D wavelet plot has neither frequency-axis tick labels nor a title; use only one title for the three plots on the right. Resolve any alignment issues caused by the colorbar appropriately. Enhance the figure aesthetically by setting all border line widths to 1 pt, using Times New Roman for all English text at 8 pt, applying appropriate decimal formatting to tick labels, and ensuring that no elements overlap. Finally, save the generated image to a complete file path suitable for Markdown insertion.}

2. Allocate node:
\myhang{2em}{Please use the Other node agent to execute the following task: load the input signal, with sampling frequency 50 Hz; perform a Fourier transform and an appropriate wavelet transform on this signal; extract and display only the 0–1 Hz frequency components for plotting. Create one single figure with total canvas width 7.06 containing: (1) a left subplot showing a 3D time-frequency plot of the wavelet transform; and (2) a right composed layout containing a 2D wavelet time-frequency plot, the original data plot placed directly above the 2D plot with aligned time axes, and the Fourier transform spectrum plot placed vertically on the left side of the 2D wavelet plot with aligned frequency axes. The original data plot must have no time-axis ticks, labels, or title. The 2D wavelet plot must have no frequency-axis ticks, labels, or title. Only one title should be used for the three plots on the right. Properly handle any misalignment caused by the colorbar on the right side of the 2D wavelet plot so that the original data plot remains aligned with the 2D plot. Choose an appropriate wavelet and beautify the figure: set all figure border line widths to 1 pt, use Times New Roman for all English text at 8 pt, use appropriate decimal formatting for axis ticks, and ensure plot elements do not overlap. Save the generated image to a complete local path suitable for Markdown insertion, and return that complete saved image path.}

3. Other node:
\myhang{2em}{The task has been completed and the result is shown in Figure \ref{fig11}.}

\vspace{1em}
\textbf{Ouput:}
\myhang{2em}{The requested analysis was performed on the input signal sampled at 50 Hz. Specifically, a Fourier transform and a wavelet-based time–frequency analysis were conducted, with visualization limited to the 0–1 Hz frequency range as requested. The final output was organized as a single composite figure with a total width of 7.06, comprising a 3D wavelet time–frequency plot on the left and, on the right, a combined layout including the original signal, a 2D wavelet time–frequency plot, and the Fourier spectrum. All specified axis alignment and formatting requirements were applied. The resulting image is presented in Figure \ref{fig11}.}
\noindent\rule{\linewidth}{1pt}

The complete dialogue for probability distribution fitting using SHM-Agents is detailed in dialogue 3.

\noindent\rule{\linewidth}{1.5pt}

\noindent{\textbf{  Dialogue 3.} Probability distribution fitting.}

\noindent\rule{\linewidth}{1pt}

\textbf{Input:}

\myhang{1em}{1.	Command:}
\myhang{2em}{Based on the input temperature data samples, use the multi-modal normal distribution as the foundation to fit a probability distribution model that can represent the temperature distribution. After the fitting is completed, draw the fitted pdf and cdf on a graph (with a width of 3.506) along with the statistical histogram of the original data. Also, mark the proportion, mean, and variance of each normal component on the graph.}
\myhang{1em}{2. Input data:}
\myhang{2em}{Temperature samples: real temperature samples from the temperature sensor. }

\vspace{1em}
\textbf{Thinking:}

\begin{figure}[!t]
	\centering
	\includegraphics[width=0.98\columnwidth]{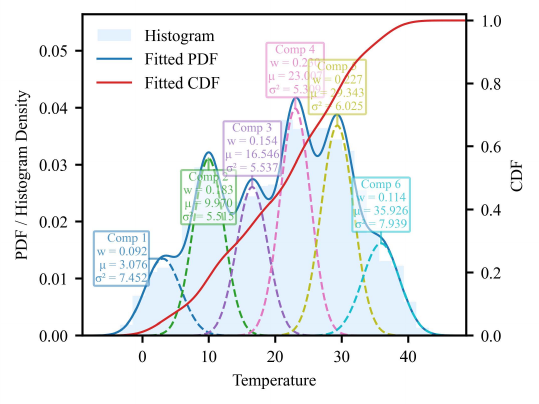}
	\caption{{Temperature probability distribution fitting graph.}}
	\label{fig12}
\end{figure}

\myhang{1em}{1. Architecture node: }
\myhang{2em}{Step 1: Use the Other node agent to fit a Gaussian mixture model based on the input temperature data determine and construct a multi-modal normal probability distribution representing the temperature distribution then draw the fitted PDF and CDF together with the statistical histogram of the original data on a figure with width `3.506` and clearly mark the proportion mean and variance of each normal component on the graph; save the generated figure to a complete local path for proper insertion.}
\myhang{1em}{ Allocate node:}
\myhang{2em}{Please use the Other node agent to load the temperature data variable, fit a Gaussian mixture model to construct a multimodal normal probability distribution that represents the temperature distribution, and then draw the fitted PDF and CDF together with the statistical histogram of the original data on one figure with width 3.506. Please clearly mark the proportion, mean, and variance of each Gaussian component on the graph, save the generated figure to a complete local path for proper insertion, and return the complete saved image path.}
\myhang{1em}{Other node:}
\myhang{2em}{The task has been completed and the result is shown in Figure \ref{fig12}.}

\vspace{1em}
\textbf{Ouput:}
\myhang{2em}{A Gaussian mixture model was fitted to the temperature data to capture its multimodal distribution. A figure with a width of 3.506 was generated to display the histogram of the original data together with the fitted probability density function (PDF) and cumulative distribution function (CDF). The proportion, mean, and variance of each Gaussian component were also annotated in the plot. The resulting image is presented in Figure \ref{fig12}.}
\noindent\rule{\linewidth}{1.5pt}

In the aforementioned simple tasks represented by wavelet transformation and probability distribution fitting, SHM-Agents are capable of autonomously generating and debugging Python code. These tasks do not require the involvement of specialized models.

\subsubsection{Modal identification}

The complete dialogue on short-term data anomaly diagnosis, data reconstruction and structural modal identification using SHM-Agents is presented in Dialogue 4.

\begin{figure*}[!t]
	\centering
	\includegraphics[width=0.98\textwidth]{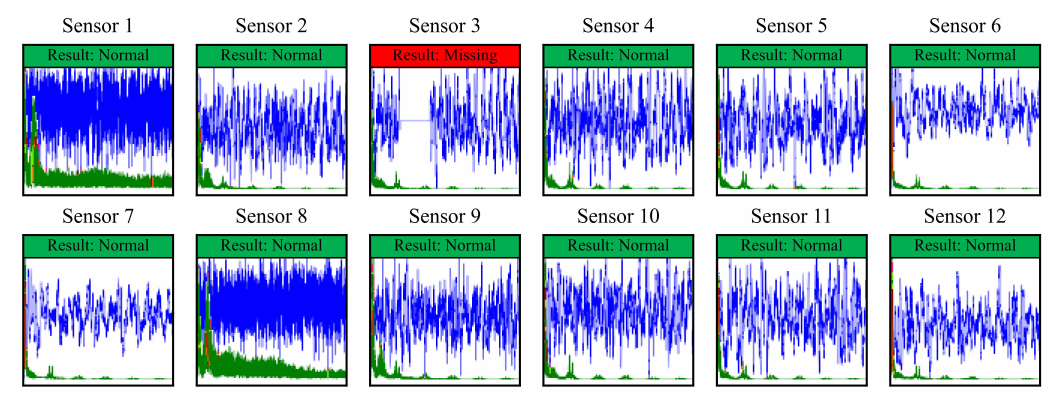}
	\caption{{ Data Anomaly Diagnosis Result. The blue lines represent the time-domain signals, while the green lines correspond to the frequency-domain signals. Above each plot, the identification results of data anomaly diagnosis are indicated.}}
	\label{fig14}
\end{figure*}

\begin{figure*}[!t]
	\centering
	\includegraphics[width=0.98\textwidth]{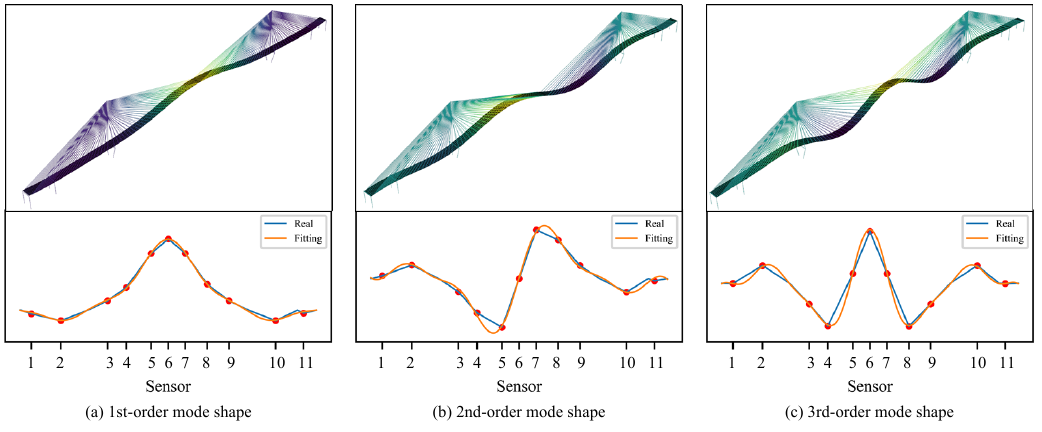}
	\caption{{ Mode shapes identification results.}}
	\label{fig15}
\end{figure*}

\noindent\rule{\linewidth}{1.5pt}

\noindent{\textbf{Dialogue 4.} structural modal identification.}

\noindent\rule{\linewidth}{1pt}

\textbf{Input:}
\myhang{1em}{1. Command:}
\myhang{2em}{Please first perform data anomaly diagnosis and data reconstruction, and then conduct structural modal identification based on the reconstructed data.}
\myhang{1em}{2. Input data:}
\myhang{2em}{Monitoring data: Monitoring data from acceleration sensors.}

\vspace{1em}
\textbf{Thinking:}

1. Architecture node: 
\myhang{2em}{Step 1: Use the Data Anomaly Diagnosis node to conduct data anomaly diagnosis on the input monitoring data \texttt{monitoring\_data}.}
\myhang{2em}{Step 2: Use the Modal identification node to identify the mode of the structure based on the input monitoring data \texttt{monitoring\_data}.}

2. Allocate node:
\myhang{2em}{Please use the Data Anomaly Diagnosis node agent to conduct data anomaly diagnosis on the input monitoring data \texttt{monitoring\_data}.}

3. Data Anomaly Diagnosis node:
\myhang{2em}{The data anomaly diagnosis has been successfully executed. The 3rd-dimensional data was identified as abnormal, with the exception type being "Missing".}
\myhang{2em}{The abnormal data has been successfully reconstructed. The reconstructed data has been saved as \texttt{Effective\_monitoring\_data}. }
\myhang{2em}{Additionally, the data anomaly diagnosis result has been captured in Figure \ref{fig14}.
}

4. Allocate node:
\myhang{2em}{Please use the Modal identification node agent to identify the mode of the structure based on the input monitoring data \texttt{Effective\_monitoring\_data}.}

5. Modal identification node:
\myhang{2em}{The static modal parameter identification has been successfully completed.
Here are the results:}
\myhang{2em}{-1st Order Frequency: 0.048, Damping Ratio: 0.001}
\myhang{2em}{-2nd Order Frequency: 0.053, Damping Ratio: 0.002}
\myhang{2em}{-3rd Order Frequency: 0.071, Damping Ratio: 0.005}
\myhang{2em}{The detailed results, including frequency data, damping ratio data, and mode shape data, have been saved locally. Figure \ref{fig15} shows the modal shape.}

\vspace{1em}
\textbf{Ouput:}
\myhang{2em}{The static modal parameter identification has been successfully completed.
Here are the results:}
\myhang{2em}{-1st Order Frequency: 0.048, Damping Ratio: 0.001}
\myhang{2em}{-2nd Order Frequency: 0.053, Damping Ratio: 0.002}
\myhang{2em}{-3rd Order Frequency: 0.071, Damping Ratio: 0.005}
\myhang{2em}{The detailed results, including frequency data, damping ratio data, and mode shape data, have been saved locally. Figure \ref{fig15} shows the modal shape.}

\noindent\rule{\linewidth}{1.5pt}

According to Dialogue 4, SHM-Agents first decomposes the structural modal identification task into two subtasks: data anomaly detection and reconstruction and structural modal identification. Initially, the data anomaly detection algorithm is used to identify abnormalities. As shown in Figure \ref{fig14}, the algorithm plots the time-domain and frequency-domain representations of each dimension of the monitoring data in a single figure. A pretrained convolutional neural network is then employed to detect anomalies in these images. After the anomalous data are reconstructed using the anomaly data reconstruction algorithm, the reconstructed data are combined with the normal data and passed to the subsequent subtask. Finally, structural modal identification is performed using the reconstructed data, and the mode-shape curves are fitted and plotted with reference to the finite element model.

The complete conversations used for analyzing changes in long-term modal parameters with SHM-Agents are detailed in Dialogue 5.

\begin{figure*}[!t]
	\centering
	\includegraphics[width=0.98\textwidth]{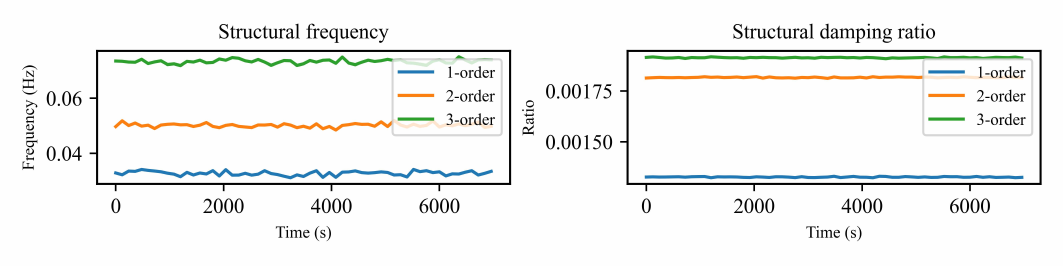}
	\caption{{Graph of changes in structural modal parameters.}}
	\label{fig13}
\end{figure*}

\noindent\rule{\linewidth}{1.5pt}

\noindent{\textbf{Dialogue 5.} Analyzing changes in long-term modal parameters.}

\noindent\rule{\linewidth}{1pt}

\textbf{Input:}
\myhang{1em}{1. Command:}
\myhang{2em}{Please first perform data anomaly diagnosis and data recovery, and then analyze the long-term variation of the structural modal parameters, with the time interval set to 1 minute.}
\myhang{1em}{2. Input data:}
\myhang{2em}{Monitoring data: Monitoring data from acceleration sensors.}

\vspace{1em}
\textbf{Thinking:}

1. Architecture node: 
\myhang{2em}{Step 1: Use the Data Anomaly Diagnosis node agent to conduct long-term data anomaly detection on the input monitoring data with a time interval set at 60 seconds.}
\myhang{2em}{Step 2: Use the Modal identification node agent to analyze the long-term changes in the structural modal parameters based on the anomaly-free monitoring data with the time interval set at 60 seconds.}

2. Allocate node:
\myhang{2em}{Please use the Data Anomaly Diagnosis node agent to conduct long-term data anomaly detection on the input \texttt{monitoring\_data} with the time interval set at 60 seconds.}

3. Data Anomaly Diagnosis node:
\myhang{2em}{The long-term data anomaly detection task has been successfully executed. Below are the results:}
\myhang{2em}{1.	Original Data Anomaly Diagnosis Result: The graph is show in Figure \ref{fig8}.}
\myhang{2em}{2.	Reconstruction Data: Abnormal data has been reconstructed successfully. The reconstructed data is stored in the variable \texttt{Effective\_monitoring\_data}.}
\myhang{2em}{3.	Reconstruction Anomaly Diagnosis Result: The graph is show in Figure \ref{fig9}.
}

4. Allocate node:
\myhang{2em}{Please use the Modal identification node agent to analyze the long-term changes in the structural modal parameters based on the anomaly-free \texttt{Effective\_monitoring\_data} with the time interval set at 60 seconds.}

5. Modal identification node:
\myhang{2em}{The analysis of the long-term changes in the structural modal parameters has been successfully completed. Please find the results in Figure \ref{fig13}.}

\vspace{1em}
\textbf{Ouput:}
\myhang{2em}{The analysis tasks requested have been completed successfully. The changes in structural modal parameters are shown in Figure \ref{fig13}.}
\noindent\rule{\linewidth}{1.5pt}

From Dialogue 5, it can be concluded that SHM-Agents performs long-term structural modal parameter analysis through a series of systematic steps. First, the Data Anomaly Diagnosis node is employed to identify anomalies and reconstruct monitoring data at predefined time intervals.  Next, the Modal identification node processes the reconstructed monitoring data to perform sequential analyses of structural modal parameters over successive time intervals. Finally, the system produces a graphical representation illustrating the variations in the first three modal frequencies and their corresponding damping ratios.

\subsubsection{Structural reliability calculation}

The complete conversations for calculating the structural reliability at the current moment and over the next 10 years using SHM-Agents are detailed in Dialogues 6 and 7.

\noindent\rule{\linewidth}{1.5pt}

\noindent{\textbf{Dialogue 6.} Structural reliability at the current moment.}

\noindent\rule{\linewidth}{1pt}

\textbf{Input:}
\myhang{1em}{1. Command:}
\myhang{2em}{Please calculate the structural reliability under the input uncertainty file.}
\myhang{1em}{2. Input data:}
\myhang{2em}{Monitoring data: Monitoring data from acceleration sensors.}

\vspace{1em}
\textbf{Thinking:}

1. Architecture node: 
\myhang{2em}{Step 1: Use the Reliability calculation node to calculate the structural reliability under the uncertainty file using the updated finite element model.}

2. Allocate node:
\myhang{2em}{Please use the Reliability calculation node agent to calculate the structural reliability under the uncertainty file \texttt{uncertainty\_file} using the updated finite element model.}

\begin{figure}[!t]
	\centering
	\includegraphics[width=0.98\columnwidth]{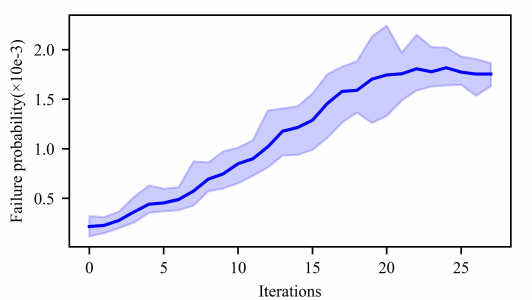}
	\caption{{ Reliability calculation process.}}
	\label{fig16}
\end{figure}

3. Reliability calculation node:
\myhang{2em}{Successfully calculated the structural reliability and failure probability under specified uncertainties, with a reliability of 99.82\%, and a failure probability of 0.18\%. The reliability index is 2.91. The reliability calculation process is shown in Figure \ref{fig16}.}

\vspace{1em}
\textbf{Ouput:}
\myhang{2em}{The structural reliability calculation has been conducted using the updated finite element model and the specified uncertainty file. The reliability of the structure has been computed at 99.82\%, with a failure probability of 0.18\%. The reliability index is 2.91. }

\noindent\rule{\linewidth}{1.5pt}

\begin{figure*}[!t]
	\centering
	\includegraphics[width=0.98\textwidth]{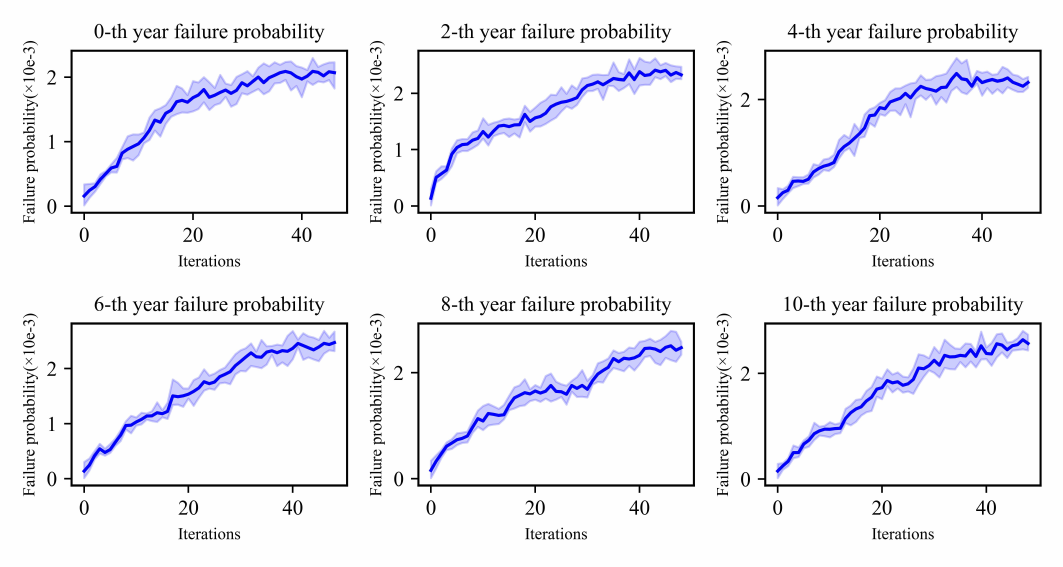}
	\caption{{The calculation curve of structural failure probability for future years.}}
	\label{fig17}
\end{figure*}

\noindent\rule{\linewidth}{1.5pt}

\noindent{\textbf{Dialogue 7.} Structural reliability over the next 10 years.}

\noindent\rule{\linewidth}{1pt}

\textbf{Input:}
\myhang{1em}{1. Command:}
\myhang{2em}{Please calculate the change in failure probability of the structure over the next 10 years under the input uncertainty file, with a time interval of 2 years.}
\myhang{1em}{2. Input data: }
\myhang{2em}{uncertainty file: Structural Uncertainty Configuration File.}

\vspace{1em}

\textbf{Thinking:}

1. Architecture node: 
\myhang{2em}{Step 1: Please calculate the change in failure probability of the structure over the next 10 years under the input uncertainty file, with a time interval of 2 years.}

2. Allocate node:
\myhang{2em}{Please use the Reliability calculation node agent to calculate the change in failure probability of the structure over the next 10 years based on the uncertainty file.}

\begin{figure}[!t]
	\centering
	\includegraphics[width=0.98\columnwidth]{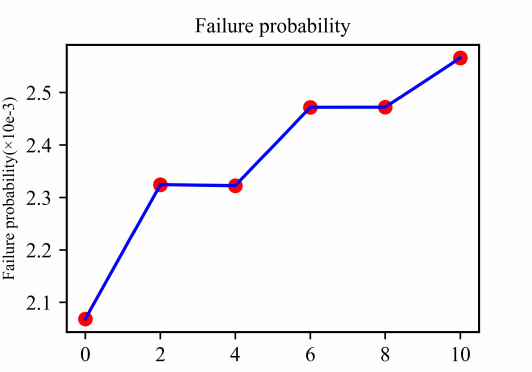}
	\caption{{The change in structural failure probability over the next 10 years, with a time interval of 2 years.}}
	\label{fig18}
\end{figure}

3.Reliability calculation node:
\myhang{2em}{Successfully calculated the change in failure probability of the structure over the next 10 years with a time interval of 2 years.
The failure probabilities for each year are shown in Figure \ref{fig17}. The changes chart of failure probability over the years is shown in Figure \ref{fig18}.
}

\vspace{1em}
\textbf{Ouput:}
\myhang{2em}{The change in failure probability of the structure over the next 10 years, using a time interval of 2 years, has been successfully calculated based on the input uncertainty file. The changes chart of failure probability over the years is shown in Figure \ref{fig18}.}

\noindent\rule{\linewidth}{1.5pt}

Dialogues 6 and 7 present structural reliability analyses for the current state and the next 10 years, respectively. In these two tasks, once the finite element model updating algorithm has been developed, the data anomaly diagnosis and reconstruction method and the structural modal identification method can be sequentially applied to obtain the actual modal parameters of the structure, as described in Dialogue 4. These parameters can then be incorporated into the finite element model updating algorithm to update the initial finite element model, thereby producing a model that more accurately reflects the actual structural condition. Based on the updated model, reliability assessment can then be performed to determine the reliability of the actual structure.

In the task of analyzing structural reliability over the next 10 years, the input uncertainty file incorporates a corrosion model describing the time-dependent reduction in the cable cross-sectional area due to environmental degradation. Driven by these physical degradation parameters, the system activates the Reliability calculation node to derive the increasing trend of the failure probability.

\subsubsection{structural fatigue damage calculation}
The complete dialogue for computing structural fatigue damage using SHM-Agents is detailed in dialogue 8.

\noindent\rule{\linewidth}{1.5pt}

\noindent{\textbf{Dialogue 8.} Computing structural fatigue damage.}


\begin{figure}[!t]
	\centering
	\includegraphics[width=0.98\columnwidth]{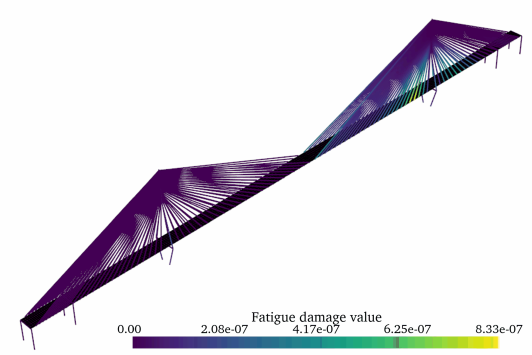}
	\caption{{Fatigue damage diagram under 30-minute vehicle loading.}}
	\label{fig20}
\end{figure}

\noindent\rule{\linewidth}{1pt}

\textbf{Input:}
\myhang{1em}{1. Command:}
\myhang{2em}{Please calculate the fatigue damage of the structure under the vehicle load for half an hour.}

\vspace{1em}
\textbf{Thinking:}

1. Architecture node: 
\myhang{2em}{Step 1: Use the Vehicle load modeling node agent to generate the vehicle load with a duration of 1800 seconds (half an hour).}
\myhang{2em}{Step 2: Use the Fatigue calculation node agent to calculate the structural fatigue damage under the generated vehicle load for the duration specified.}

2.Allocate node:
\myhang{2em}{Please use the Vehicle load modeling node agent to generate the vehicle load with a duration of 30 minutes. }

3. Vehicle load modeling node:
\myhang{2em}{The vehicle load with a duration of 30 minutes has been successfully generated. The results are stored in a Python dictionary named \texttt{V\_load}}

4. Allocate node:
\myhang{2em}{Please use the Fatigue calculation node agent to calculate the structural fatigue damage under the generated vehicle load \texttt{V\_load} for half an hour.}

5. Fatigue calculation node:
\myhang{2em}{The fatigue damage calculation was successfully completed. The fatigue damage results are shown in Figure \ref{fig20}, and are saved in data  \texttt{Fatigue\_damage\_calculation}.}

\vspace{1em}
\textbf{Ouput:}
\myhang{2em}{The fatigue damage calculation was successfully completed. The fatigue damage results are shown in Figure \ref{fig20}. }
\noindent\rule{\linewidth}{1.5pt}

In Dialogue 8, SHM-Agents segments the fatigue damage calculation into two distinct subtasks: vehicle load generation and fatigue damage calculation. The detailed execution process is as follows: SHM-Agents utilizes the vehicle load generation algorithm to sample from a pre-established vehicle load model, obtaining vehicle load data for a duration of 30 minutes. Following this, the fatigue damage calculation algorithm is applied to determine the fatigue damage on the structure due to the vehicle load, with the damage results being presented in Figure \ref{fig20}.

\subsubsection{Structural response calculation}

The complete dialogue for computing structural response using SHM-Agents is detailed in dialogue 9.

\begin{figure}[!t]
	\centering
	\includegraphics[width=0.98\columnwidth]{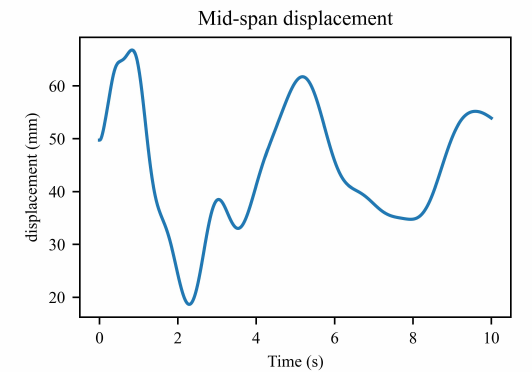}
	\caption{{Mid-span displacement diagram under 10s vehicle loading.}}
	\label{fig21}
\end{figure}

\noindent\rule{\linewidth}{1.5pt}
\textbf{Dialogue 9.} Computing structural response.

\noindent\rule{\linewidth}{1pt}

\textbf{Input:}
\myhang{1em}{1. Command:}
\myhang{2em}{Please calculate the mid-span displacement of the structure under a 10-second vehicle load.}

\vspace{1em}
\textbf{Thinking:}

1. Architecture node: 
\myhang{2em}{Step 1: Use the Vehicle load modeling node to generate the vehicle load with a duration of 10s.}
\myhang{2em}{Step 2: Use the Response calculation node to calculate the structural response under the generated vehicle load.}

2.Allocate node:
\myhang{2em}{Please use the Vehicle load modeling node agent to generate the vehicle load with a duration of 10 seconds. }

3. Vehicle load modeling node:
\myhang{2em}{The vehicle load with a duration of 10 seconds has been successfully generated. The results are stored in a Python dictionary named \texttt{V\_load}}

4. Allocate node:
\myhang{2em}{Please use the Response calculation node agent to calculate the structural response under the generated vehicle load \texttt{V\_load}.}

5. Fatigue calculation node:
\myhang{2em}{Finite element load solving has been successfully completed. The mid-span displacement diagram has been successfully drawn and the image is showed in Figure \ref{fig21}.}

\vspace{1em}
\textbf{Ouput:}
\myhang{2em}{The structural response under the 10-second vehicle load was successfully calculated using the Response calculation node agent. The mid-span displacement diagram has been successfully drawn and the image is showed in Figure \ref{fig21}}
\noindent\rule{\linewidth}{1.5pt}

SHM-Agents employs the same approach used in Dialogue 9 to complete structural response calculations, partitioned into two distinct subtasks: vehicle load generation and structural response calculation. The execution processes of the first subtasks align entirely with those in Dialogue 8, except for variations in vehicle load sampling time. Upon sampling the vehicle load, SHM-Agents activates the response calculation algorithm, utilizing Pyansys to compute the structural response under the vehicle load. Subsequently, it extracts and plots the structure's mid-span displacement diagram, as illustrated in Figure \ref{fig21}.

\subsubsection{Bridge knowledge question and answer}

The complete dialogue for bridge knowledge Q\&A using SHM-Agents is detailed in dialogue 10.

\noindent\rule{\linewidth}{1.5pt}
\textbf{Dialogue 10.} Bridge knowledge Q\&A.

\noindent\rule{\linewidth}{1pt}

\textbf{Input:}
\myhang{1em}{1. Command:}
\myhang{2em}{Please inquire about the materials used for the bridge's stay cables.}

\vspace{1em}
\textbf{Thinking:}

1. Architecture node: 
\myhang{2em}{Step 1: Use the RAG node agent to inquire about the materials used for the bridge's stay cables.}

2.Allocate node:
\myhang{2em}{Please use the RAG node agent to inquire about the materials used for the bridge's stay cables.}

3. RAG node:
\myhang{2em}{The stay cables of the bridge are usually made of low-relaxation high-strength prestressed steel strands. These steel strands comply with the regulations of "Steel Strand for Prestressed Concrete" (GB/T 5224-2003). The diameter of each single steel strand is 15.2 millimeters, the cross-sectional area of the steel strands is 140 square millimeters, the standard strength is 1860 MPa, and the elastic modulus is $1.95 \times 10^{5}\ \mathrm{MPa}$.}

\vspace{1em}
\textbf{Ouput:}
\myhang{2em}{The materials used for the bridge's stay cables typically include low-relaxation, high-strength pre-stressed steel strands. These steel strands comply with the specification 'GB/T 5224-2003' for pre-stressed concrete. Each steel strand has a diameter of 15.2 mm, a strand area of 140 square mm, a standard strength of 1860 MPa, and an elastic modulus of $1.95 \times 10^{5}\ \mathrm{MPa}$.}
\noindent\rule{\linewidth}{1.5pt}

As demonstrated in Dialogue 10, in the bridge knowledge question-answering task, the SHM-Agents directly retrieves relevant information from the knowledge database based on the input question and then integrates this information to answer the user's inquiry.

The results from the aforementioned dialogues demonstrate that SHM-Agents effectively completed a variety of integrated tasks in SHM autonomously, significantly enhancing the overall efficiency.

\section{Limitations}
The limitations of SHM-Agents can be summarized as follows:
\begin{enumerate}
    \item SHM-Agents' problem-solving abilities are directly correlated with the capabilities of LLM chat models. Specifically, when employing gpt-4o as the LLM chat model, the operational accuracy of the system surpasses 95\% but is not 100\%.

    \item In the context of SHM tasks, SHM-Agents demonstrate accuracy and efficiency comparable to various specialized algorithms. Overall, while SHM-Agents offer a platform for efficient algorithm execution, the ultimate results are contingent upon the performance of the specialized algorithms themselves.
    
	
\end{enumerate}

\section{Conclusion}

This paper introduces SHM-Agents, a generalist–specialist integrated agent system designed to efficiently perform both individual and composite tasks in SHM, such as reliability assessment and fatigue estimation. 

SHM-Agents adopts a modular architecture consisting of process agent nodes, skill agent nodes and global modules. Process agent nodes are responsible for planning execution workflows and allocating tasks, while skill agent nodes perform specific SHM functions. The global modules oversee information management throughout system operation. The coordinated interaction among these three components enables the agent system to accomplish SHM tasks accurately and efficiently.

An application example involving an actual cable-stayed bridge demonstrates that SHM-Agents can correctly and efficiently complete a variety of SHM tasks. SHM-Agents effectively formulated task workflows and accurately assigned tasks to each skill agent, thereby ensuring the accuracy of variable transmission throughout the process and maintaining a clear task execution procedure.

It should be noted that several planned specialized algorithms have yet to be implemented, including those for damage identification, finite element model updating, visual surface damage detection, special events assessment, and digital twin–based simulation and inference. However, our agent-based system architecture allows newly developed agents to be continuously added, which represents a significant advantage.

\section{Declaration of conflicting interests}

The author(s) declared no potential conflicts of interest with respect to the research, authorship, and/or publication of this article.

\section{Funding}
This research was supported by grants from the National Natural Science Foundation of China
(Grant No. 52425804, U23A20660).

\bibliographystyle{SageH} 

\end{document}